# Developer Insights into Designing AI-Based Computer Perception Tools

Maya Guhan[1], Meghan E. Hurley[1], Eric A. Storch[2], John Herrington[3], Casey Zampella[3], Julia Parish-Morris[3], Gabriel Lázaro-Muñoz[4], Kristin Kostick-Quenet[1]

[1]Center for Ethics and Health Policy, Baylor College of Medicine, Houston, TX
[2]Department of Psychiatry and Behavioral Sciences, Baylor College of Medicine, Houston, TX
[3]Department of Child and Adolescent Psychiatry and Behavioral Sciences, Children's Hospital of Philadelphia, Philadelphia, PA
[4]Center for Bioethics, Harvard Medical School, Boston, MA

## Abstract

Artificial intelligence (AI)-based computer perception (CP) technologies use mobile sensors to collect behavioral and physiological data for clinical decision-making. These tools can reshape how clinical knowledge is generated and interpreted. However, effective integration of these tools into clinical workflows depends on how developers balance clinical utility with user acceptability and trustworthiness.

Our study presents findings from 20 in-depth interviews with developers of AI-based CP tools. Interviews were transcribed and inductive, thematic analysis was performed to identify 4 key design priorities: 1) to account for context and ensure explainability for both patients and clinicians; 2) align tools with existing clinical workflows; 3) appropriately customize to relevant stakeholders for usability and acceptability; and 4) push the boundaries of innovation while aligning with established paradigms.

Our findings highlight that developers view themselves as not merely technical architects but also ethical stewards, designing tools that are both acceptable by users and epistemically responsible (prioritizing objectivity and pushing clinical knowledge forward). We offer the following suggestions to help achieve this balance: documenting how design choices around customization are made, defining limits for customization choices, transparently conveying information about outputs, and investing in user training. Achieving these goals will require interdisciplinary collaboration between developers, clinicians, and ethicists.

**Introduction**

The integration of artificial intelligence (AI)-based computer perception (CP) technologies into clinical decision-making represents a profound transformation in how medical knowledge is generated, interpreted, and applied. The term "computer perception" stems from the term "computer vision" but acknowledges a broader set of sensor modalities beyond vision alone (e.g., "hearing" through microphones; detecting motion via accelerometry; gauging physiological responses like glucose levels, heart rate or neural activity through various sensors).[1] The suite of technologies that employ CP approaches include digital phenotyping, computational behavioral analysis, computational psychiatry, and other data-driven systems that analyze continuous behavioral data passively collected from patients outside of clinical and lab settings using wearable, mobile or implantable devices to capture a wide spectrum of "ethological" data, i.e., behaviors occurring in their naturalistic contexts. Data streams can include activity and GPS tracking, physiological biometrics such as heart rate/blood pressure, vocal acoustics, call and text logs and social media usage, as well as mobile phone-based surveys ("ecological momentary assessments") among many others. Unlike previous paradigms in mental and behavioral health interventions that rely primarily on validated clinical assessments administered by clinicians or self-reports from patients, CP systems enable algorithmic inferences about a person's mental or behavioral states that rely more directly on computer, rather than human, perception.

These CP technologies have the potential to augment clinician expertise, improve diagnostic accuracy, and streamline healthcare workflows.[2, 3] While most of these approaches remain in early stages of development and validation, their integration into care is imminent.[4-6] However, their potential impacts on clinical decision-making and quality of care are not fully understood. In addition to concerns around accuracy and reliability (trustworthiness), CP technologies raise ethical questions around privacy, overreliance and other impacts on the patient-clinician relationship, regulation and governance, and technological reductionism or solutionism.[7-12] Developers of these tools are tasked with designing systems that balance technical precision, usability, and ethical safeguards, to ensure that they offer valid, reliable and clinically meaningful insights. How developers conceptualize the role of their tools and respond to the informational needs of their users influences the decisions they make about how their systems synthesize data, what information is emphasized, and how recommendations align with clinical reasoning.[13, 14] These decisions have critical implications for the integrity and clinical utility of CP tools, which will increasingly be used to curate and synthesize clinical information, a role historically reserved for human experts. This shift in the responsibility for curating medical insights raises critical questions about how CP technologies select and present information, the extent to which clinicians are provided with essential tools and information to critically evaluate these insights, and how the content of AI-driven recommendations can introduce unintended biases or constraints on clinical reasoning.[15-17]

Experts and regulators alike have called for greater attention to the practical and ethical considerations influencing developers' decisions about how to design their algorithms and convey their outputs in user interfaces.[18, 19] To better understand the rationale, as well as the ethical and practical implications, behind systems design for CP tools, it is essential to understand what features developers consider to be integrally important for CP tools to be usefully and responsibly employed by physicians and patients. These considerations are crucial for ensuring that AI remains an augmentative force in clinical reasoning rather than an opaque system that constrains medical judgement.[17]

This paper investigates developers' perspectives and understandings about what features should characterize AI-based CP tools to enable end-users (in this case, primarily clinicians and patients) to use them effectively and responsibly. We present empirical insights from in-depth interviews with developers actively involved in advancing CP technologies. Our findings contribute to emerging knowledge about how AI systems are being conceptualized and designed with certain end-user needs in mind. These insights, in turn, open pathways for dialogue with research and clinical communities to address the clinical, practical and ethical implications of developers' content and design choices, to collaboratively and optimally shape CP systems in line with user needs.

**Methods**

We conducted in-depth, semi-structured interviews with developers of AI-based computer perception tools (n=20) to explore their perspectives on potential benefits, risks and concerns around the integration of CP technologies into clinical care, and particularly, what considerations factor into their algorithmic and interface design choices.

<u>Participants</u>. Participants included developers of AI-based CP tools designed to detect biomarkers (facial, gesture, or other behavioral indicators) of at least one of a broad range of mental health and/or neurodegenerative conditions, including autism, Tourette Syndrome, anxiety and depressive disorders, obsessive-compulsive disorder, and Attention Deficit/Hyperactivity Disorder (ADHD), among others. Participants were identified through online literature search and through existing professional networks and interviewed between January 2023 and August 2023.

<u>Data Collection</u>. Interview guide questions were organized by specific domains and constructs, including: perceived benefits and concerns regarding integrating CP tools into clinical care, impacts on care, perceived ethical considerations for automatic and passive detection of emotional and behavioral states, perceived accuracy and potential for misinterpretation/-attribution/-classification of symptoms or conditions, clinical utility and actionability, data security and privacy concerns, potential for unintended uses, perceived generalizability and potential for bias. These domains were chosen based on issues raised in the clinical and ethics literatures and with the guidance of experienced bioethicists and mental health experts. Initial drafts of the interview guides were piloted with two psychologists (ES, CJZ) specializing in adolescent mental health, resulting in minor clarifications in wording. Interviews were conducted via a secure video conferencing platform (Zoom for Healthcare) and lasted an average of ~45 minutes. This study was approved by the Baylor College of Medicine Institutional Review Board (H-52227), which waived a requirement for written consent; thus, participants provided verbal consent.

<u>Data Analysis.</u> Interviews were audio-recorded, transcribed verbatim, and analyzed using MAXQDA software. Led by a qualitative methods expert (KK-Q), team members developed a codebook to identify thematic patterns in participant responses to questions addressing the topics above. Each interview was coded by merging work from two separate coders (MG, MH) to reduce interpretability bias and enhance reliability. We used Thematic Content Analysis to inductively identify themes by progressively abstracting relevant quotes, a process that entails reading every quotation to which a given code was attributed, paraphrasing each quotation (primary abstraction), further identifying which constructs were addressed by each quotation (secondary abstraction), and organizing constructs into themes.[20] To enhance the validity of our findings, all abstractions were validated by at least one other member of the research team. In rare cases where abstractions reflected different interpretations, members of our research team met to reach consensus.

**Table 1. Demographics for Interviewed Developers**

| Developers (n=21) | | n = | % Total |
|---|---|---|---|
| Gender | Male | 18 | 86% |
| | Female | 3 | 14% |
| Role | Clinician-Developer | 4 | 19% |
| | Developer | 17 | 81% |
| Domain | Industry | 15 | 71% |
| | Academic | 3 | 14% |
| | Cross-Sector | 3 | 14% |

**Results**

Our findings reveal consensus among developers that AI-driven CP tools hold strong potential to transform clinical decision-making by enhancing how data are collected, analyzed, visualized, and utilized by healthcare providers. Specifically, developers highlighted CP's ability to predict future clinical trends, integrate multimodal analysis for deeper insights, and detect subtle patterns that may escape human observation. They unanimously considered these advancements to offer promising new avenues for optimizing clinician decision-making and improving patient outcomes. However, developers also identified four key considerations to ensure that AI-based computer perception tools and outputs are used ethically and effectively. These include: (1) Output Context and Explainability; (2) Alignment with Existing Clinical Workflows and Paradigms; (3) Optimal (but not Over-) Customization; (4) Balancing Innovation and Responsibility. These themes are discussed in turn below, with illustrative quotations presented in Table 2.

**Theme 1. Output Context and Explainability**

Developers viewed clinicians as playing a crucial role in interpreting and translating CP-generated data into actionable patient care decisions. Developers highlighted the fact that the data must be presented in ways that accommodate different stakeholders' perspectives and levels of statistical or computational expertise. They pointed out that this requires developers to format outputs to offer both high-level summaries for quick, digestible insights (e.g., for patient end-users, or even clinicians with limited technical expertise or for use in fast-paced clinical settings) as well as more detailed information and recommendations for complex case evaluations and evidence-based decision making. In addition, respondents said that patients need to receive insights commensurate with their literacy level, and that do not generate unnecessary anxiety.

Many developers also felt that, because clinicians are the primary gatekeepers of AI-driven recommendations, ensuring interpretability, transparency and explainability in outputs is particularly important because it enables clinicians to not only evaluate the outputs but also to convey their logic and rationale to patients, as well as their role in the clinician's recommendations. Developers considered it essential to interpret and contextualize outputs within the broader context of a patient's care and thus advocated for platforms that enable visualization of longitudinal trends in patient symptom expressions, as well as integration with other key aspects of a patient's symptom and treatment history. They also felt that outputs should be accompanied by certain technical information that might allow clinicians (and patients) to understand which observations played a significant role in system inferences, a feature which could facilitate interpretability, explainability, and communication with patients. Finally, many argued that outputs should be directly relevant and specific to a patient's care and thus lead to actionable treatment options.

**Theme 2. Alignment with Existing Workflows and Paradigms**

One of the most significant challenges in AI adoption is its integration into existing clinical workflows. Developers recognized that for AI tools to gain widespread acceptance, they must enhance clinical decision-making, provide clear time- and cost-saving benefits, and integrate seamlessly into electronic health records (EHRs) and hospital information technology (IT) systems. Developers emphasized that CP insights also need to align with physicians' existing paradigms, expectations and decision-making needs to be actionable. Outputs that are perceived to fall too far outside of the scope of clinicians' existing procedures or expectations are anticipated to generate distrust or skepticism.

While they unanimously felt that CP systems offer the potential for enhanced diagnostic precision, predictive analytics, and improved decision-support mechanisms, developers stated that their integration into clinical practice is contingent upon how clinicians interpret outputs and how these tools fit within existing decision-making frameworks. Developers highlighted that AI tools should be designed with iterative feedback loops, ensuring that developers refine system functionalities in response to clinician and patient needs as well as develop appropriate educational programs to teach end users about the outputs of these algorithms, including their intended value for clinical decision making and empirically informed suggestions for interpreting and integrating them into care.

Developers thus considered clinicians and patients to be key partners in the development of CP tools and felt a responsibility to be responsive to their perspectives, not only to ensure utility but also to facilitate their adoption and effectiveness in practice. Developers recognized that CP systems must be co-designed with clinicians to ensure that its outputs are relevant, interpretable, and seamlessly integrated into workflows that complement existing workflows and do not add cognitive burden to intended end users.

**Theme 3. Optimal (but not Over-) Customization**

Many developers were enthusiastic about the potentials of customization for enhancing utility and acceptability, including tailoring the content, format, timing or other output interface features to user preferences. They noted that ongoing research and development of CP tools mirrors trends towards personalization and customization in the broader field of AI, following the rationale that these systems will lead to greater relevance of outputs and motivate engagement and acceptability among users. However, some viewed customization as a double-edged sword.

Developers acknowledged that some degree of customization enhances end user engagement by allowing CP systems to align with existing informational preferences, paradigms and workflows (as discussed under Theme 2). However, at least one developer pointed out that over-customization could introduce bias, reduce system reliability, and hinder broader applicability across diverse healthcare settings, raising practical and ethical questions about how much customization is appropriate before it begins to constrain clinical reasoning. Developers emphasized that balancing tailored outputs with access to a wide range of clinical insights, including those that may push the boundaries of user preference, is essential for maintaining responsible, evidence-based decision-making with CP.

**Theme 4. Balancing Innovation with Responsibility**

Developers raised ethical and practical concerns about balancing alignment with innovation. Many said they are keenly aware that for their technologies to be adopted in clinical practice, they must align with clinicians' current workflows, cognitive models, and diagnostic reasoning frameworks (Theme 1). However, many developers said they are also simultaneously trying to create tools that go beyond what clinicians already know or are used to seeing – tools that are capable of shifting paradigms by redefining how disease is measured, categorized, and understood. As one developer put it, the goal is not to merely confirm what the clinician suspects, but to reveal insights that would not have been visible through conventional diagnostics. Thus, this introduces a critical tension: the more novel the insight, the more likely it is to be met with skepticism or rejection.

Developers thus explained that they find themselves in a continual process of negotiation. On one hand, they feel the need to design tools that clinicians, patients and other potential stakeholders will use and that map onto well-known clinical nosologies and understandings. On the other hand, they recognize that the true promise of CP tools lies in their ability to challenge those categories and to surface patterns or risks that escape even human attention. They also recognize that clinicians, as key mediators of CP outputs, bear the burden of interpreting, validating, and communicating these novel insights to patients in ways that users will be receptive to. As such, many developers said they face an ethical responsibility to design systems that users find to be appealing and useful (for example, assisting clinicians in their existing workflows, or helping patients to more effectively monitor and manage their health) but that also challenge them to consider novel or alternative perspectives that expand their established understandings about illness (e.g., viewing certain conditions as dimensional or dynamic rather than discreet or static). Developers acknowledged that, to bridge the potential gap between developers' intentions and end user expectations, there must be continuous collaboration between these stakeholder groups with each informing the other about their intentions and rationale in advocating for certain features of CP systems.

**Table 2**

| Theme | Illustrative Quotes |
|---|---|
| **Need for Context and Explainability** | Usually it's one of the clinicians who reviews it (the system's output)… They have a relationship with them (the patient). The **clinician can really put it in context and make it meaningful** to not only their past, but the current state of the individual as well as future. Then **that is really powerful, when the data aren't just the data standing alone on a piece of paper, it's actually interpreted and thought about within the context of their care**" (Developer 6).<br><br>I think where my overarching philosophy, and the company's philosophy around this is that **whatever data you show to a patient** that may be related to their disease, it **should be actionable in a clear way**. **It shouldn't be a mystery what to do with it**. It shouldn't send them down a rabbit hole of Googling." (Developer 9).<br><br>"I think **an algorithm shouldn't just tell you a recommendation, it should give you a lot of meta information around that, why, explainability,** which is obviously a hot topic. **It should tell you the rationale,** if possible, and it **should allow the user to interrogate the algorithm regarding what variables it used, what variables are making the most contribution to the prediction or recommendation**. If possible, if there was **some sort of certainty, index or confidence index** of the algorithms; The algorithm not only can inform you of what it thinks, let's say, or what it recommends, it can provide **a level of uncertainty or certainty around that**" (Developer 15). |
| **Align with Existing Clinical Workflows and Paradigms** | "So [**clinicians] will say, "Well, we have this particular need**." Most of the time, it's an actual problem. They come up with a problem. They say, "Oh, we have this problem. We are wondering." So, obviously if they're using the tool fast, the conversation is easier because they already have an idea the tool might be able to address that. "So we were wondering, can your tool be able to address this?"<br>**And so in a very responsible way,** the way we do it is you first go through some sort of an interview, one hour, two hours, **first understand the problem very deep**. What is the problem? What are they trying to accomplish? Who's going to use it? How are they going to use it? How were they going to access it? And **then we figure out where we are going to place it in a workflow**." (Developer 17).<br><br>"So I think one key feature is **customization to a specific user seems to really improve implementation or dissemination of these models**, which is interesting. The other thing is time savings or scale. So if you could do something not quite as good as a human, but you **save them large amounts of time**, they're very likely to implement it. Whereas if you **do the same thing with really high performance and it costs them in their clinical workflow, they're unlikely to use it, which is interesting**. …Those are not things that |
| | physicians often articulate. They'll often verbalize these very high performance metrics, but I would say **similarity to individual end user's practice and integration within workflows matters a lot**…I think that captures that experience more than anything else, but **does this fit in my workflow?** When I look around, **do people I respect and value their opinions, are they using it? Does it make me seem like I'm achieving relevant benchmarks** for my own criteria for success? I think those are the kinds of things you see, but **the thing that they don't articulate is this customizability thing**. And then **workflow time savings are huge**" (Developer 19). |
| **Optimal (but not Over-) Customization** | "We perhaps **should have different information designs or data visualizations for different stakeholders**. We're in the area where everybody talks with dashboards and so on, but perhaps you should be able to **interrogate data at different levels of detail,** and we should consider the intuitiveness of different data visualizations and **which data visualization is appropriate** for, as you say, patients" (Developer 15).<br><br>"So I think **if you ask physicians what they want, they often want you to approximate whatever their practice is**. Which **may not necessarily be more accurate**, which is a little weird. So physicians will say, "Oh, I don't like that. I would've done that different." But you can sometimes point and say, "Yeah, but **your segmentation or your diagnosis in fact is outside of the distribution of other human observers." And they'll say, "Yeah, I don't care. I want it to look like me**" (Developer 19). |
| **Balance Responsibility with Innovation** | "**Clinicians need new, different kinds of data to do a better job. Researchers need to work with clinicians to figure out what that data will look like and this is a partnership**. We need to serve clinicians in the future by collecting data that they don't even know that they will want versions of later" (Developer 7).<br>"I think from a clinician perspective, they **need to empower the clinician in some way.** And empowerment here, it might be, "Now, I used to see five patients, now I'm seeing 10. So, I've been able to **improve my productivity** in that sense." Or the other direction is, "I've been **able to cut costs**. And so, it means that I used to need 10 people to do the job now and need only five of them." Because the downstream, I think if we look at that cumulatively, that is some sort of it's obviously **going to improve the system in one way or the other**" (Developer 17). |

## Discussion

*The Innovation-Implementation Paradox*

    Our findings reveal that CP developers face a previously unappreciated paradox stemming from their felt need to create paradigm-shifting innovations, and the potentially competing need to align these innovations with existing paradigms to ensure stakeholder acceptability, trust and uptake. Prior research on AI trustworthiness confirms that users expect AI tools to be transparent, interpretable, and aligned with evidence-based medicine to ensure responsible integration into

clinical workflows.[21, 22] Clinicians serve as critical mediators and translators of CP-generated insights, a role that carries significant ethical and legal implications. Developers acknowledge that while CP can enhance decision-making, the ultimate responsibility for making clinical recommendations remains with clinicians, making transparency and explainability central concerns.[23] In line with this, our results indicate that a strong priority for developers is to ensure that clinicians receive CP-generated information in ways that facilitate end user interpretation and understanding. They suggested a range of design options, including "layered" information that gives users the choice to receive easily digested summaries of patient trends or inferences that enable quick decision-making, as well as the capacity to expand these summaries to access deeper and more technical insights (e.g., explanatory variables, validation criteria, confidence intervals) when preferable. Developers offered the rationale that outputs must be interpretable at multiple levels of expertise, ensuring that different types of users, such as specialists compared to general practitioners, can engage with them meaningfully. They also pointed out that these layered outputs could also help to mitigate liability risks by reducing interpretive ambiguity and promoting critical evaluation of CP outputs when integrating them into clinical decisions. They noted that these efforts should be complemented with focused training for clinicians to promote effective and responsible interpretation of CP-derived insights without overwhelming them with technical complexities, particularly in cases where outputs may involve uncertainty.

There is widespread consensus among the researcher and developer communities that design principles like those described above are critical to supporting transparency, trustworthiness, and accountability of AI-driven systems.[24] Ethics and legal scholars have likewise advocated for the inclusion of AI "nutrition facts labels" that clearly communicate system performance metrics, as well as "datasheets for data sets" that convey features such as data set composition, imbalance or other information relevant to evaluating their quality and generalizability.[25, 26] However, beyond implementing these informational features that are now widely endorsed, developers must make decisions about system features that are preference-sensitive. These decisions involve wide-ranging considerations about content, communication, visualization, and workflow integration that are often specific to certain use cases and anticipated user populations. Developers of CP systems are no exception, as they attempt to anticipate the preferences of their envisioned end users: primarily clinical professionals in mental health and related disciplines, and/or patients who are the intended consumers of their CP tools and platforms. Our developer interviewees appeared to be prepared and even eager to respond to user preferences from these groups.

However, a tension arises from an impression that developers are creating tools that many of these envisioned end users are not yet ready for, as they do not always share the paradigms required to fully appreciate and capitalize on their tools. Clinician end users have likely received training in diagnostic frameworks outlined by the Diagnostic and Statistical Manual of Mental Disorders (DSM) or the International Classification of Diseases (ICD) and may be expecting outputs to align with these frameworks. Patient consumers are likely to conceptualize mental health concerns in similar terms. But CP tools gained momentum in response to calls by renowned mental health experts to *reconsider* established ways of conceptualizing and addressing mental illness.[27] They advocate a shift from thinking about illness in ontological terms (discrete diagnostic categories) towards viewing symptoms as manifestations of more dimensional, dynamic and continuous disease states. The resulting new paradigm – the Research Domains Criteria (RDoC) framework – lent new purposes to remote sensing devices that were just beginning to appear on the market (e.g., in the form of consumer wearables and smart phones) and gave life to approaches like digital phenotyping and computational psychiatry.[28] These paradigms highlighted the strong potential of leveraging algorithms to identify patterned symptom constellations that would enable better therapeutic targeting and outcome prediction. Developers in our study shared a view of CP technologies as helping to advance this revolutionary vision of mental healthcare and expressed a desire to create tools that enable stakeholders to explore these innovative approaches. Simultaneously, they recognized the importance of providing tools that generate insights that stakeholders can make immediate sense of in the context of existing paradigms, and that assist individuals of varying levels of expertise and technological orientation.[29] Doing so requires that developers straddle a dual obligation to create tools that push paradigms into new frontiers while also meaningfully aligning with existing paradigms.

*The Customization Challenge*

These findings underscore the complex ethical and trade-offs involved in aligning AI systems with clinical needs. Developers discussed "customization" as the primary site of this ethical tension, which they described in a range of forms – from adapting tools to local workflows and institutional values, to tailoring interfaces to individual clinician preferences. Developers shared that each of these approaches holds significant potential to improve usability and foster trust. However, developers also raised concerns that certain forms of customization – particularly those tied to individual preferences – could inadvertently compromise objectivity and reinforce existing biases. These concerns are underscored by recent scholarship highlighting that customization can restrict access to information that may be critical for quality decision making but may not readily conform to user informational preferences or habits.[17] Over-customization, as described by several developers in our study, occurs when AI systems are adapted so closely to individual clinician preferences, historical usage patterns, or institutional norms that they begin to suppress disconfirming or unfamiliar information. In such cases, personalization no longer functions to enhance interpretability or workflow alignment but rather reinforces entrenched heuristics and limits cognitive diversity. Curating information to align with user preferences without considering what information users *should* know (according to developers' own values, orientations, or to broader stakeholder consensus and/or reasonable identification of objective ground truths) could reinforce cognitive biases and limit exposure to critical information.[17] Thus, a key challenge for developers, one with both practical and ethical implications, is deciding how much CP tools should be adapted to fit clinician preferences without compromising their ability to provide unbiased, objective insights into patient functioning.

As an example, some end users may want CP tools to analyze biological and behavioral markers to identify the presence or severity of a DSM-defined illness. While some developers are excited about such possibilities, others remain skeptical that the digital markers central to CP systems can ever meaningfully map onto existing diagnostic categories. They feel that digital markers are better conceptualized as data points within broader constellations of data (i.e., digital phenotypes) that fall along various spectra, in line with RDoC or related frameworks (e.g., the Hierarchical Taxonomy of Psychopathology, or "HiTOP" model).[30] For these developers, it may feel disingenuous or even futile to customize algorithmic tools for end users with fundamentally different expectations or intentions. Nevertheless, they recognize that such preferences must be considered if they are to design CP tools with significant acceptability and uptake.

Thus, several developers have argued that customization should be carefully designed to enhance workflow integration rather than simply cater to individual clinician preferences.[31]

*Recommendations for Responsible Customization and Integration*

These challenges reflect a broader need for collaborative, interdisciplinary engagement between developers, clinicians, and ethicists. Developers highlight the importance of iterative feedback loops, ensuring that CP tools evolve in response to clinician needs while maintaining a focus on their originally intended capacities and utilities. These feedback loops could take many forms, including:

<u>Document Design Choices and Rationale</u>. Developers could benefit from systematically documenting how design choices around customization are made and justified, particularly with regard to how CP systems identify classification thresholds for diagnostic inferences, or filter, rank, or highlight other clinical insights.[17] Making these prioritization strategies transparent – through clear developer documentation or embedded system notes – would allow both clinical users and researchers to better evaluate how customization may influence patient outcomes over time.

<u>Define the Limits and Justifications for Customization</u>. Developers should incorporate mechanisms that define the limits of allowable customization – ensuring that while systems are responsive to different clinical environments, they do not lose generalizability or drift away from standards of care.[29] This ensures that AI maintains its usability and acceptability without compromising its intended utility. Further, developers could implement tiered customization options backed by explicit paradigmatic justifications, allowing user choice and flexibility while preserving core features that enable objective pattern seeking in patient data.

<u>Build-in Transparency, Explainability, and Confidence</u>. To enhance transparency and clinician trust, CP systems should also present clear indicators of confidence or reliability, as well as (potentially expandable, hover, or drop-down) explanations of how outputs were derived. Providing structured rationales for AI-driven recommendations can help clinicians to critically evaluate AI outputs.

<u>Invest in User Literacy and Training</u>. Developers should not be solely responsible for ensuring AI or CP literacy among end users. However, consistent with other assessment tools which offer detailed user manuals, model cards, or other instructions for use, developers could offer structured educational materials ensuring that clinicians understand the intended utility, strengths, and limitations of CP systems. Additionally, user-centered decision-support dashboards should be developed to ensure that AI outputs are intuitively presented and interpretable. These measures can contribute to the effective integration of CP into care; however, they are unlikely to assist clinicians in translating CP insights to patients, a topic that remains in need of future, focused research. While developers in our study largely conceptualized end users to be clinicians, this group is likely (and already does, in many cases involving direct-to-patient or -consumer feedback) to include patients and even caregivers, guardians, or other surrogate decision makers. How best to tailor these technologies to these groups in ways that encourage receptivity, acceptability, and understanding remains understudied.

## Conclusion

As AI-based computer perception tools move toward greater integration into healthcare, developers face complex and often competing demands: to innovate while ensuring usability, to challenge clinical paradigms while aligning with them, and to customize while preserving objectivity. This study highlights that developers are not merely technical architects but also ethical stewards, tasked with designing tools that are both clinically actionable and epistemically responsible. Through layered outputs, structured transparency, and thoughtful customization, CP systems can support clinicians in making informed, context-sensitive decisions—without becoming either rigid confirmation engines or indecipherable black boxes. Achieving this balance will require sustained collaboration among developers, clinicians, patients, and ethicists. Only through such interdisciplinary partnerships can we ensure that CP tools fulfill their transformative promise while upholding accepted standards of care and trustworthiness. Future research must continue to explore how design decisions reflect and shape epistemological assumptions in clinical practice and how CP technologies can be designed to support ethical and human-centered innovation.

## Conflict of interest

ES reports receiving research funding to his institution from the Ream Foundation, International OCD Foundation, and NIH. He was formerly a consultant for Brainsway and Biohaven Pharmaceuticals in the past 12°months. He owns stock less than $5000 in NView. He receives book royalties from Elsevier, Wiley, Oxford, American Psychological Association, Guildford, Springer, Routledge, and Jessica Kingsley. The remaining authors declare that the research was conducted in the absence of any commercial or financial relationships that could be construed as a potential conflict of interest.

## Acknowledgements

This study was funded by National Center for Advancing Translational Sciences (grant number: 1R01TR004243) and National Institute of Mental Health (grant number: 3R01MH125958).

ChatGPT was used in the creation of an initial outline of this paper, which was substantively modified by co-authors.